# Specially designed $B_4C/SnO_2$ nanocomposite for photocatalysis: traditional ceramic with unique properties

*Paviter Singh[1], Gurpreet Kaur[1], Kulwinder Singh[1], Bikramjeet Singh[1], Manpreet Kaur[1], Manjot Kaur[1], Unni Krishnan[1], Manjeet Kumar[2], Rajni Bala[3], Akshay Kumar\*[1]*

[1]Advanced Functional Materials Lab., Department of Nanotechnology, Sri Guru Granth Sahib World University, Fatehgarh Sahib- 140 407, Punjab, India
[2] Department of Electrical Engineering, Incheon National University, 406772, Incheon, South Korea
[3] Department of Mathematics, Punjabi University, Patiala- 147 002, Punjab, India
Corresponding author E-mail: akshaykumar.tiet@gmail.com

## Abstract

Boron carbide: A traditional ceramic material shows unique properties when explored in nano-range. Specially designed boron based nanocomposite has been synthesized by reflux method. The addition of $SnO_2$ in base matrix increase the defect states in boron carbide and shows unique catalytic properties. The calculated texture coefficient and Nelson Riley factor shows that the synthesized nanocomposite have very high defect states. Also this composite is explored for the first time for catalysis degradation of industrial used dyes. The industrial pollutants such as Novacron red and methylene blue dye degradation analysis reveal that the composite is an efficient catalyst. Degradation study shows that 1 g/L catalyst concentration of $B_4C/SnO_2$ degrade Novacron red Huntsman dye upto 97.38% approximately in 20 minutes under sunlight irradiation time. This water insoluble catalyst can be recovered and reused.

## Introduction

Dyes and pigments are extensively used in textile industries, food, cosmetics, paper, leather and plastic industries for coloring the products. The estimated production of different dyes is about approximately 0.7 to 7 million tons per year [1-3]. The extent of pollution caused by discharge of dyes into the environment is unknown. The use of dyes is main source of environmental pollution and also become a big threat to aquatic life. The removal of dye from waste water is crucial need of time for better sustainability as most of the dyes and their secondary products are carcinogenic or mutagenic and noxious in nature [4 -5]. The various methods are to be used by researchers to decolourize different dyeing effluents. The industrial waste water treatment involves processes such as physical (adsorption) [6-7], chemical (ozonation) [8], biological [9-11] reverse osmosis and as well as photocatalysis [12]. The high cost physical as well as chemical methods have failed to treat waste water as this economically unrealistic and also results in unavoidable secondary pollution. The photocatalysis technology is economically viable method used for degradation of dyes at ambient conditions. Carbides have attracted attention of researchers in recent years due to its extraordinary properties such as high mechanical strength, high melting point and their chemical inertness owing to their potential applications in thermionic electron sources. Nanostructured carbides have been used in various fields such as biomaterials, light weight/high strength materials, high temperature resistant materials, semi-conducting devices [13-14]. Various synthesis methods have been used in synthesis of boron carbide nanostructures such as carbothermal method from reduction of boron oxide ($B_2O_3$) over 1000 °C, thermal decomposition method, gaseous reaction between boron trichloride ($BCl_3$) and a methane hydrogen mixture in presence of radio frequency argon plasma, reduction of $BCl_3$ by $CH_4$ at 1500 °C with laser [15-16]. A wide range of high temperature synthesis methods can be used for preparation of boron carbide nanostructures directly from boron and carbon [17]. But these methods are economically not viable due to the use of expensive precursors. As metal oxides has attracted considerable attraction in research field due to their physical and chemical properties. Among them, tin oxide ($SnO_2$) has a wide band gap (3.6 eV) and high thermal stability so $SnO_2$ is used as a potential candidate as a photocatalyst [18-19]. The barriers in photocatalytic efficiency, in case of $SnO_2$ nanoparticles are their aggregation as size decreases and electron hole recombination process. $B_4C/SnO_2$ composite has been synthesized using wet chemical synthesis method in order to obtain the improved photocatalytic efficiency for waste water treatment. The transition metal oxides, phosphides, sulfides replaced the high cost noble metals based electrocatalysts as well as photo-catalysts in the past years due to their low cost and high activity [20]. Though, corrosion and passivation under acidic conditions cause main hurdle for most of these materials. Besides, there is need of developing a stable and catalytically active material for photocatalysis process in order to reduce water pollution. The transition metal oxides usually showed their failures in field of active site engineering. In recent years, the catalysts that contain non-metallic nature and earth abundance such as carbon are employed as alternative catalyst materials for water purification process. Also, $B_4C$ is a semiconductor with band gap of about 1.5 eV [21].

In this work, we emphasized on the synthesis of an efficient catalyst $B_4C/SnO_2$ composite for removal of industrial pollutants with the purpose of water purification process. The $B_4C/SnO_2$ composite has been synthesized using reflux method. The composite of $SnO_2$ with $B_4C$ as base matrix can show remarkable photocatalytic properties. Also the catalyst can be recovered and reused. Present study deals with synthesis and photocatalytic properties analysis of $B_4C/SnO_2$ nanocomposite. Industrial pollutants methylene blue (MB) and Novacron red Huntsman (NRH) dyes were used as target materials. Their degradation analysis is studied in details and degradation mechanism is also proposed for this study.

## Experimental Section

*Materials and methods*

Boric acid ($H_3BO_3$, 99.9%), activated magnesium (Mg, 98%) and acetone (used as carbon source), $SnCl_2$, Hydrochloric acid were purchased from Sigma Aldrich. All the chemicals were used as received without any further purification.

*Synthesis of $B_4C/SnO_2$ catalyst*

$B_4C$ nanoparticles were successfully synthesized using solvothermal method [33-34]. The $B_4C/SnO_2$ photocatalyst was prepared by reflux method. The freshly prepared aqueous solutions of $SnCl_2$, $B_4C$ and HCl were added to magnetically stirred round bottom flask respectively and refluxed at 100 °C for 5 hours. The obtained product was cooled to room temperature naturally. The as prepared sample was collected and washed with distilled water so that neutral pH is obtained. The washed precipitates were collected and dried in vacuum at 80 °C for 6 hour.

*Characterization*

The dried powder of the $B_4C/SnO_2$ composite was characterized by powder X-ray diffraction (XRD). The XRD pattern with diffraction intensity versus 2θ was recorded in a Rigaku instrument with Cu-Kα radiation (λ=1.5418 Å). Transmission Electron Microscope (TEM) was carried out on TECNAI G2 20 FEI at 200 keV in order to study the morphology of synthesized material. Optical absorption spectrum was studied using UV-visible Shimadzu UV-2600 spectrophotometer.

*Photocatalysis Experiment*

The photocatalytic activities of $B_4C/SnO_2$ (1 g/L) were evaluated by degradation of aqueous solutions of methylene blue (MB) dye and a textile dye Novacron red Huntsman (NRH) (1 mg/L). All experiments were carried out at room temperature. The aqueous solutions were magnetically stirred for 30 minutes in dark to get the adsorption desorption equilibrium followed by sunlight irradiation. The maximum absorption wavelength of MB at 664 nm was observed. Typically, 20 mg of photocatalyst (1g/L) was added into 20 mL of 1 mg/L MB and NRH aqueous solution. Analytical samples were taken from reaction systems after specified time period and centrifuged to separate photocatalysts before analysis. The concentration of $B_4C/SnO_2$ photocatalyst was varied from 0 g/L to 1 g/L. The changes in absorptional intensity in spectra with different catalyst dosage were studied using UV-visible spectrophotometer.

## Result and Discussion

*XRD analysis*

XRD pattern of the synthesized $B_4C/SnO_2$ composite is shown in figure 1 (a). The collected pattern was compared with $B_4C$ and $SnO_2$ JCPDS cards i.e. 35-0798 and 41-1445 respectively. The different diffraction peaks arise from $B_4C$ and $SnO_2$ respectively. The pattern confirms the formation of $B_4C$ and $SnO_2$ phase in the synthesized sample. The broadening of diffraction peaks were used for the determination of crystallite size. The crystallite size was calculated using Debye Scherer formula [22]. The calculated average crystallite size of the synthesized material is equal to ~26 nm.

Further, the XRD pattern was used to determine the texture coefficient for the synthesized composite. The texture coefficient provides the information about the preferred growth orientation of the material. Higher the value of texture coefficient deviated from unit value; more will the growth. For the calculation of texture coefficient, standard intensities related to the diffraction planes were taken from standard JCPDS cards (35-0798 and 41-1445). Texture coefficient [23-24] is calculated using the following relationship:

$$TC(h_ik_il_i) = \frac{I(h_ik_il_i)}{I_0(h_ik_il_i)} \left[\frac{1}{n}\sum_{i=1}^{n} \frac{I(h_ik_il_i)}{I_0(h_ik_il_i)}\right]^{-1} \qquad (1)$$

where TC*(hkl)*, I*(hkl)* and $I_0$*(hkl)* are texture coefficient of the plane specified by miller indices, specimen and standard intensities (taken from JCPDS cards) respectively for a given diffraction peak. The value of n represents the number of different peaks. The texture coefficient analysis reveals that the synthesized material is more grown along (211) with a texture coefficient value 4.5131. The growth of the synthesized composite along (211) can be ascribed to the presence of defects in the synthesized sample. Further, the presence of defect states has also been studied from XRD results. The variation of Δd/d with Nelson-Riley factor [25-26] determined from XRD pattern shown in figure 1 (b). The standard d-spacing values of $B_4C$ were taken from ICDD card no. 35-0798. It is observed that in case of composite, the scattered-ness of $\Delta d/d$ values increased. The scattered-ness of $\Delta d/d$ values indicate the density of defect states/stacking faults present in the samples [27]. Therefore, it can be concluded that the $SnO_2$ incorporation leads to an increase in the defect states/stacking fault density. The effect of these higher defects states/stacking faults in the dye degradation have been discussed at the end of this section.

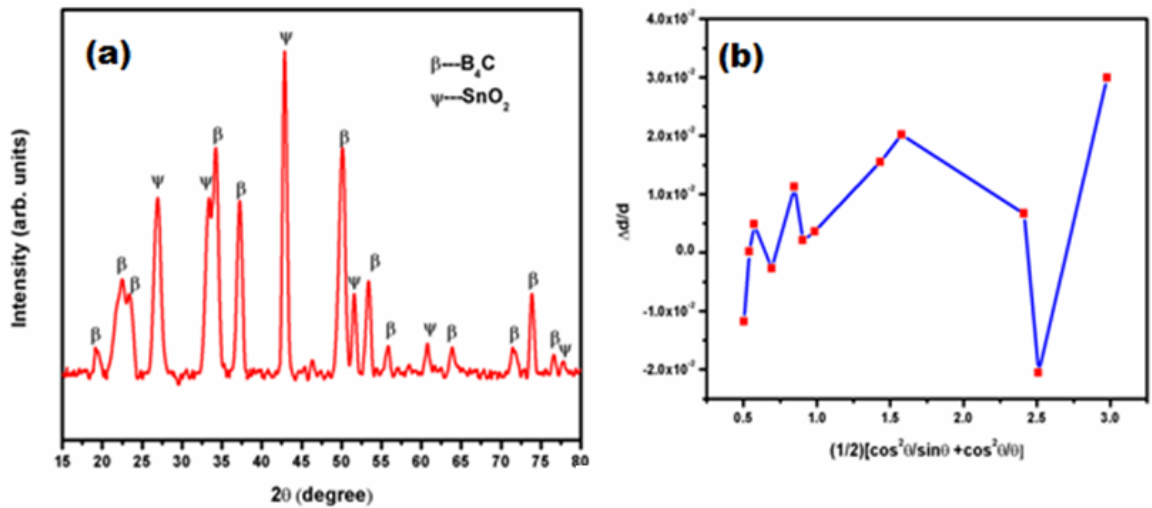

Figure 1 (a) XRD pattern of as synthesized $B_4C/SnO_2$ sample (b) Plot of $\Delta d/d$ versus Nelson-Riley factor

*TEM analysis*

TEM image of the as synthesized $B_4C/SnO_2$ is shown in figure 2. TEM image revealed that the average size of the synthesized particles is ~30 nm and are spherical in nature. The agglomeration of particles can be ascribed to small sizes and weak quantum confinement.

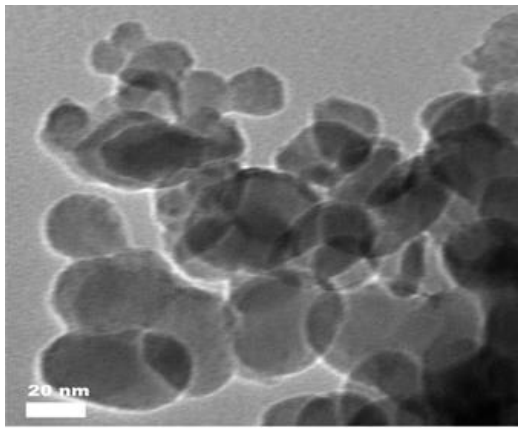

Figure 2 TEM image of synthesized $B_4C/SnO_2$ nanostructures

*Mechanism followed*

The presence of structural defects and distortion in $B_4C/SnO_2$ influence its structure. These inherent structural defects results in $B_4C/SnO_2$ with high efficiency in sunlight harvesting and makes it a good catalyst for industrial pollutants. The existence of defects causes the downshift in conduction band and available the new mid gap states that enable the boron carbide as visible light harvesting material [28]. The defects have also shown their impact on carrier relaxation dynamics, results in charge separation by trapping electron and holes [29-31]. As Nelson riley plot as well as texture coefficient indicates the presence of structural defects in $B_4C/SnO_2$. $SnO_2$ can absorb UV wavelength light and results in electron hole separation. $SnO_2$ makes electron availability to conduction band of boron carbide. These electrons and holes help to produce the $^{\bullet}OH$ radical and results in degradation of dyes.

*Photocatalysis analysis*

As the synthesized $B_4C/SnO_2$ composite, employed as a photocatalyst for the degradation of MB and NRH in water under sunlight irradiation. The effect of concentration loadings of $B_4C/SnO_2$ catalyst on degradation of used dyes (details of MB and NRH are shown in table 1) was studied. The degradation efficiency of a dye was calculated from change in the concentration of MB and NRH dye using following formula:

$$\text{Degradation ratio \%} = \frac{C_0 - C_t}{C_0} \times 100 \qquad (2)$$

where $C_0$ and $C_t$ were concentrations of dye before and after that reaction.

Figure 3 (a-d) and 4 (a-d) shows the absorption intensity changes of both the dyes with different concentration of catalyst (0 g/L to 1 g/L) under sunlight irradiation for homogeneous time interval which specifies the degradation of dye.

**Table 1. Chemical structure and characteristics of methylene blue dye and a textile dye Novacron red Huntsman**

| Chemical name | Methylene Blue | Novacron red Huntsmen |
|---|---|---|
| Chemical formula | $C_{16}H_{18}ClN_3S$ | $C_{18}H_{14}N_2Na_2O_8S_2$ |
| Molecular Weight (g/mol) | 319.85 | ~ 496.46 g/mol |
| $\lambda_{max}$ (nm) | 664 | 524 |
| Abbreviated Name | MB | NRH |
| Chemical structure | 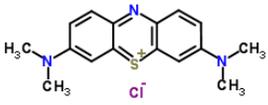 | 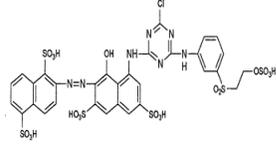 |

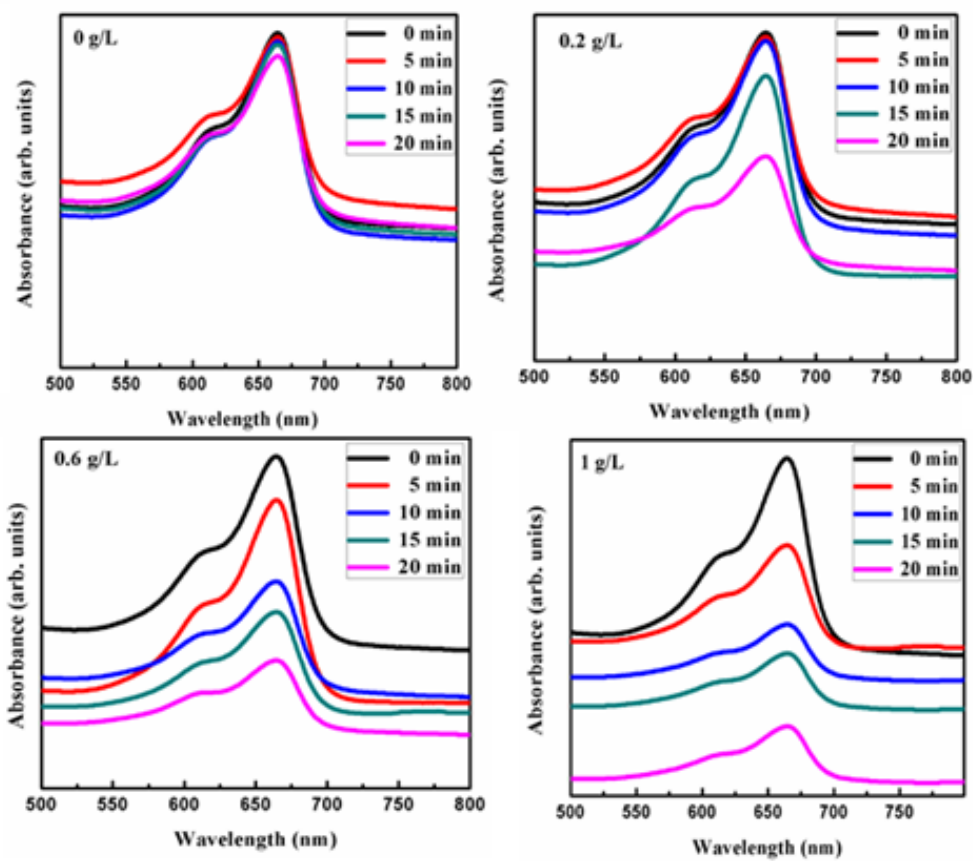

Figure 3 (a-d) Photodegradation plot of organic dye methylene blue dye (1 mg/L) aqueous solution after continuous sunlight irradiation in presence of $B_4C/SnO_2$ catalyst loading (a) 0g/L (b) 0.2 g/L (c) 0.6 g/L (d) 1 g/L

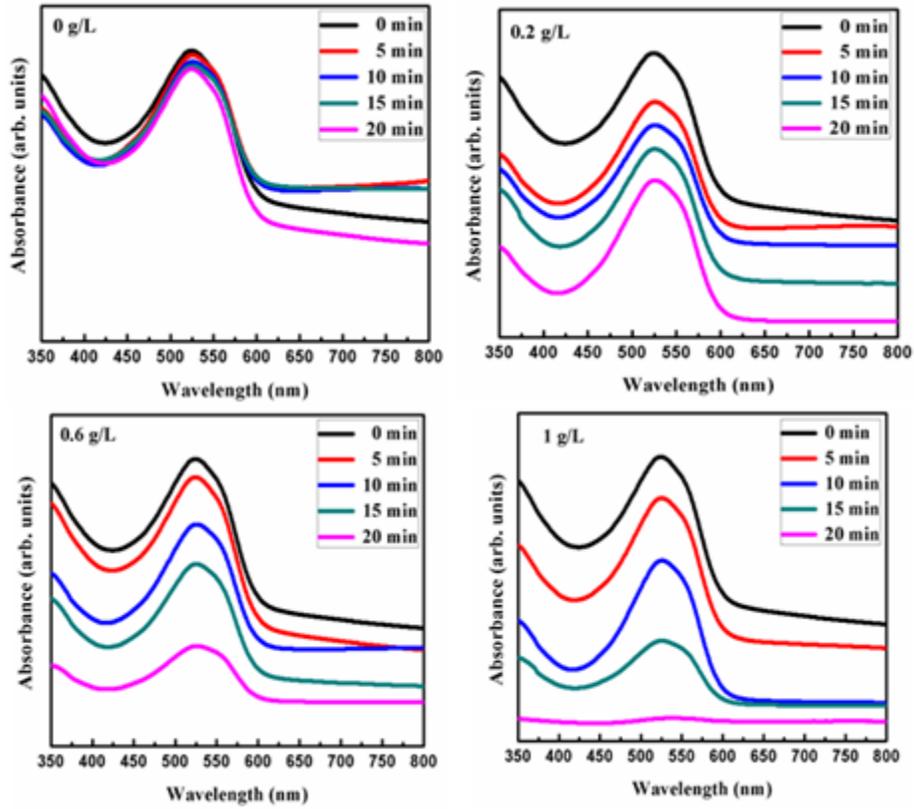

Figure 4 (a-d) Photodegradation plot of textile dye Novacron red dye (1 mg/L) aqueous solution after continuous sunlight irradiation in presence of $B_4C/SnO_2$ catalyst loading (a) 0 g/L (b) 0.2 g/L (c) 0.6 g/L (d) 1 g/L

The removal efficiency of MB and NRH with different concentration of $B_4C/SnO_2$ catalyst was described in table 2. The average decolourization rate was calculated for different catalyst loading (figure 5b) for methylene blue dye in a bar graph presentation mode using following relation:

$$\text{Average degradation rate} = \frac{C \times D\% \times 1000}{100 \times t} \quad (3)$$

where C and D% are initial concentration of dye solution and degraded dye after time t respectively. The average degradation rate of methylene blue dye with 1 g/L $B_4C/SnO_2$ catalyst is calculated as 5.0425 % (approximately) (figure 5b).

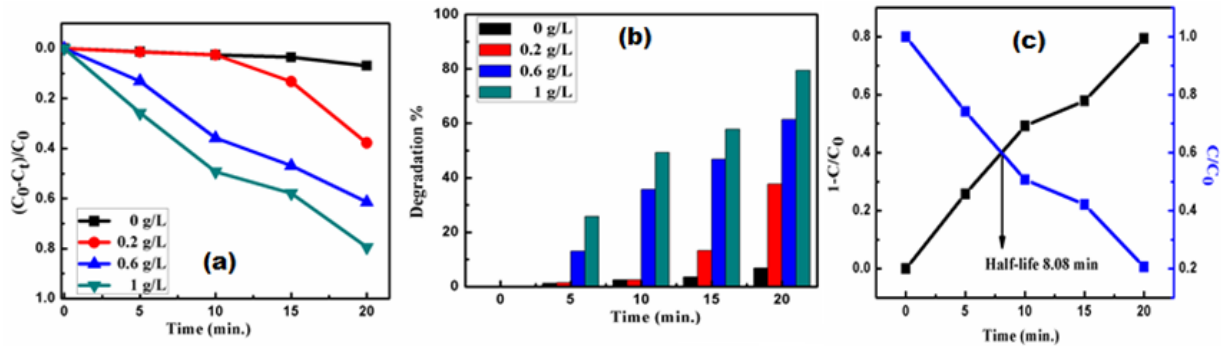

Figure 5 (a) Degradation efficiency of $B_4C/SnO_2$ catalyst with aqueous dispersion of methylene blue (MB) dye (1mg/L) as function of irradiation time (b) Average degradation rate of different concentrations of catalyst $B_4C/SnO_2$ for methylene blue (MB) at uniform time interval under sunlight irradiation (c) Half-life dye estimation curve from $C/C_0$ and degradation efficiency $1-C/C_0$ for 1 g/L of $B_4C/SnO_2$ catalyst loading in methylene blue dye (1 mg/L)

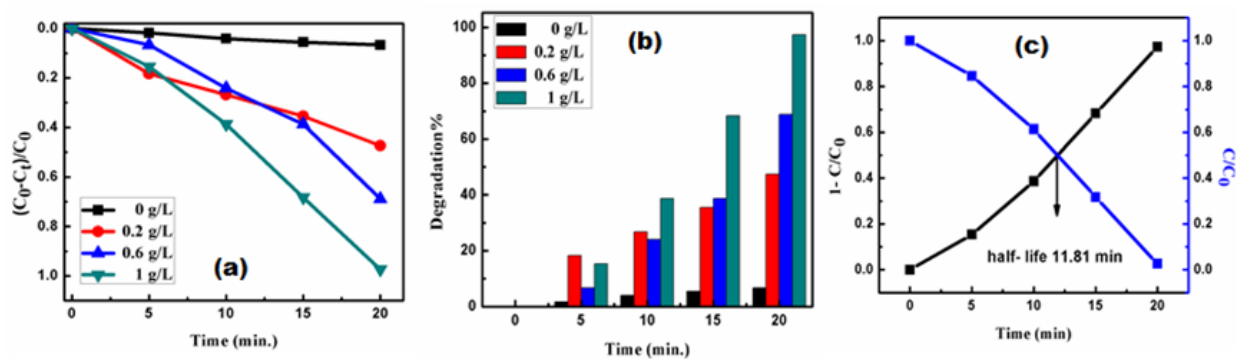

Figure 6 (a) Photocatalytic degradation efficiency of different concentrations of $B_4C/SnO_2$ catalyst loading to Novacron red (1mg/L) (b) Average degradation rate of different concentration of catalyst $B_4C/SnO_2$ for Novacron red Huntsman dye (NRH) at uniform time interval under sunlight irradiation (c) Half-life dye estimation curve from $C/C_0$ and degradation efficiency $1-C/C_0$ for 1 g/L of $B_4C/SnO_2$ catalyst loading in Novacron red dye (1 mg/L)

Figure 3 (a-d) shows the change in absorption intensities for the MB dye (1 mg/L) aqueous solution in presence of different concentrations of $B_4C/SnO_2$ catalyst under sunlight irradiation in uniform time interval of 20 minutes. Methylene blue has absorption peak at 664 nm and a shoulder at 610 nm. Both absorption bands intensity get reduced as time of sunlight irradiation increases. This leads to the decomposition or degradation of chromo-phoric group of dye into simple intermediate molecules with smaller molecular size.

Figure 6 represents the degradation efficiency of $B_4C/SnO_2$ catalyst for degradation of MB dye. The methylene blue degradation efficiency is enhanced approximately up-to 79.41 % for 1g/L $B_4C/SnO_2$ catalyst under sunlight than other concentrations of $B_4C/SnO_2$ photocatalyst. The half-life of dye defined as time at which concentration of dye became half which was calculated using intersection curve of $C/C_0$ (MB concentration) and $1-C/C_0$ (degradation efficiency) that was estimated approximately as 8.08 minutes (figure 5c).

**Table 2. Degradation ratio of methylene blue dye and a textile dye Novacron red (1 mg/L) with different catalyst dose of $B_4C/SnO_2$ respectively**

| Dye | Concentration of $B_4C/SnO_2$ catalyst loading (g/L) | Degradation efficiency (%) |
|---|---|---|
| MB | 1 | 79.41 |
|  | 0.6 | 61.42 |
|  | 0.2 | 37.76 |
|  | 0 | 6.969 |
| NRH dye | 1 | 97.38 |
|  | 0.6 | 68.89 |
|  | 0.2 | 47.38 |
|  | 0 | 6.686 |

The photo-degradation kinetics of NRH dye (1mgL$^{-1}$) with different concentrations of $B_4C/SnO_2$ catalyst under sunlight irradiation was also studied. The NRH dye has a strong absorption band at 524 nm as shown in figure 4. The absorption peak intensity declined as time of sunlight irradiation increased from 0 to 20 minutes (figure 4a-d). The degradation efficiency is estimated to be approximately 97.38 % (figure 6a) i.e. an excellent performance of $B_4C/SnO_2$ catalyst under sunlight irradiation. The average degradation rate for different concentration of $B_4C/SnO_2$ catalyst for Novacron red Huntsman dye was displayed as bar graph representation (figure 6b). The average degradation rate of Novacron red dye was calculated (using equation 4) as 43.82% (figure 6b). As the concentration of catalyst increased, the decrease in half life was observed [32].The half-life of NRH dye was estimated as 11.81 minutes (figure 6c).

## Conclusion

The $B_4C/SnO_2$ composite synthesized by simple reflux method has been employed as catalyst for degradation of an organic and a textile dye i.e. MB and NRH under sunlight irradiation. The composite acts as an efficient photocatalyst due to the presence of defect states for the removal of industrial pollutants that are noxious to the humans as well as marine life. The effect of concentration of composite as catalyst on degradation under sunlight irradiation was studied. For 1 g/L $B_4C/SnO_2$ catalyst, degradation efficiency of about 79.41 % was achieved with MB dye under sunlight irradiation in 20 minutes. The catalyst (1 g/L) competently degrades the NRH dye (1mg/L) with 97.38 %. The unique catalytic properties of $B_4C/SnO_2$ make it an alternative material in field of photocatalysis.

## Acknowledgements

This work was funded by Board of Research in Nuclear Sciences, Department of Atomic Energy (DAE), India under project no. 34/14/41/2014-BRNS. This work was supported by DST project No. EMR/2016/002815.